\documentclass{ws-ijmpe}
\usepackage{graphicx}

\newcommand{\bc}{\begin{center}}
\newcommand{\ec}{\end{center}}
\newcommand{\be}{\begin{equation}}
\newcommand{\ee}{\end{equation}}
\newcommand{\bea}{\begin{eqnarray}}
\newcommand{\eea}{\end{eqnarray}}

\begin{document}

\markboth{Sch\"afer}{From Trapped Atoms to Liberated Quarks}

\catchline{}{}{}{}{}
%%%%%%%%%%%%%%%%%%%%%%%%%%%%%%%%%%%%%%%%%%%%%%%%%%%%%%%%%%%%%%%%%%%%

\title{From Trapped Atoms to Liberated Quarks}

\author{\footnotesize Thomas Sch\"afer}

\address{Department of Physics, North Carolina State University,\\
Raleigh, NC 27695\\
thomas\_schaefer@ncsu.edu}

\maketitle

\begin{history}
%\received{(received date)}
%\revised{(revised date)}
%\accepted{(Day Month Year)}
%\comby{(xxxxxxxxxx)}
\end{history}

\begin{abstract}
We discuss some aspects of cold atomic gases in the unitarity limit
that are of interest in connection with the physics of dense hadronic
matter. We consider, in particular, the equation of state at zero 
temperature, the magnitude of the pairing gap, and the phase diagram
at non-zero polarization.
\end{abstract}

%%%%%%%%%%%%%%%%%%%%%%%%%%%%%%%%%%%%%%%%%%%%%%%%%%%%%%%%%%%%%%%%%%%%
\section{Introduction}
\label{sec_intro}
%%%%%%%%%%%%%%%%%%%%%%%%%%%%%%%%%%%%%%%%%%%%%%%%%%%%%%%%%%%%%%%%%%%%

 The QCD phase diagram contains several strongly interacting quantum 
gases or liquids. Experiments at the Relativistic Heavy Ion Collider 
(RHIC) indicate that at temperatures close to the critical temperature 
$T_c$ the quark gluon plasma is strongly interacting. The strongly 
interacting plasma (sQGP) is characterized by a very small viscosity 
to entropy ratio and by its large opacity for high energy jets 
\cite{rhic:2005}. In the opposite regime of large baryon density and 
small temperature the phase diagram features several strongly interacting 
quantum liquids. Electrically neutral matter at very low density contains 
mostly neutrons. Dilute neutron matter has positive pressure and is stable 
even at very low density. However, the neutron-neutron scattering length 
is very large and as a consequence neutron matter is strongly correlated 
even if the density is low \cite{Friedman:1981qw}. At densities on the 
order of several times nuclear matter density hadronic matter undergoes 
a phase transition to color superconducting quark matter. At extremely 
large density this phase is weakly coupled, but at densities near 
the transition to nuclear matter the gap and the critical 
temperature are probably large \cite{Rapp:1999qa}. 

 There are many questions about strongly interacting phases of QCD
that have attracted a lot of interest. Some of these questions are: 
What are the relevant degrees of freedom? Are quasi-particle pictures
appropriate? What are the transport properties? Is it true that there
is a universal bound on the viscosity, and is this bound saturated
in any of the strongly interacting phases of QCD?

 In this contribution we shall study a simpler system in which many 
of the same questions can be addressed. We consider a cold, dilute 
gas of fermionic atoms in which the scattering length $a$ of the 
atoms can be controlled experimentally. These systems can be realized 
in the laboratory using Feshbach resonances, see \cite{Regal:2005}
for a review. A small negative scattering length corresponds to a
weak attractive interaction between the atoms. This case is known
as the BCS limit. As the strength of the interaction increases the
scattering length becomes larger. It diverges at the point where 
a bound state is formed. The point $a=\infty$ is called the 
unitarity limit, since the scattering cross section saturates the 
$s$-wave unitarity bound $\sigma=4\pi/k^2$. On the other side of 
the resonance the scattering length is positive. In the BEC limit 
the interaction is strongly attractive and the fermions form deeply 
bound molecules.

%%%%%%%%%%%%%%%%%%%%%%%%%%%%%%%%%%%%%%%%%%%%%%%%%%%%%%%%%%%%%%%%%%%%
\begin{figure}
\begin{minipage}{0.49\hsize}
\bc\includegraphics[width=0.95\hsize]{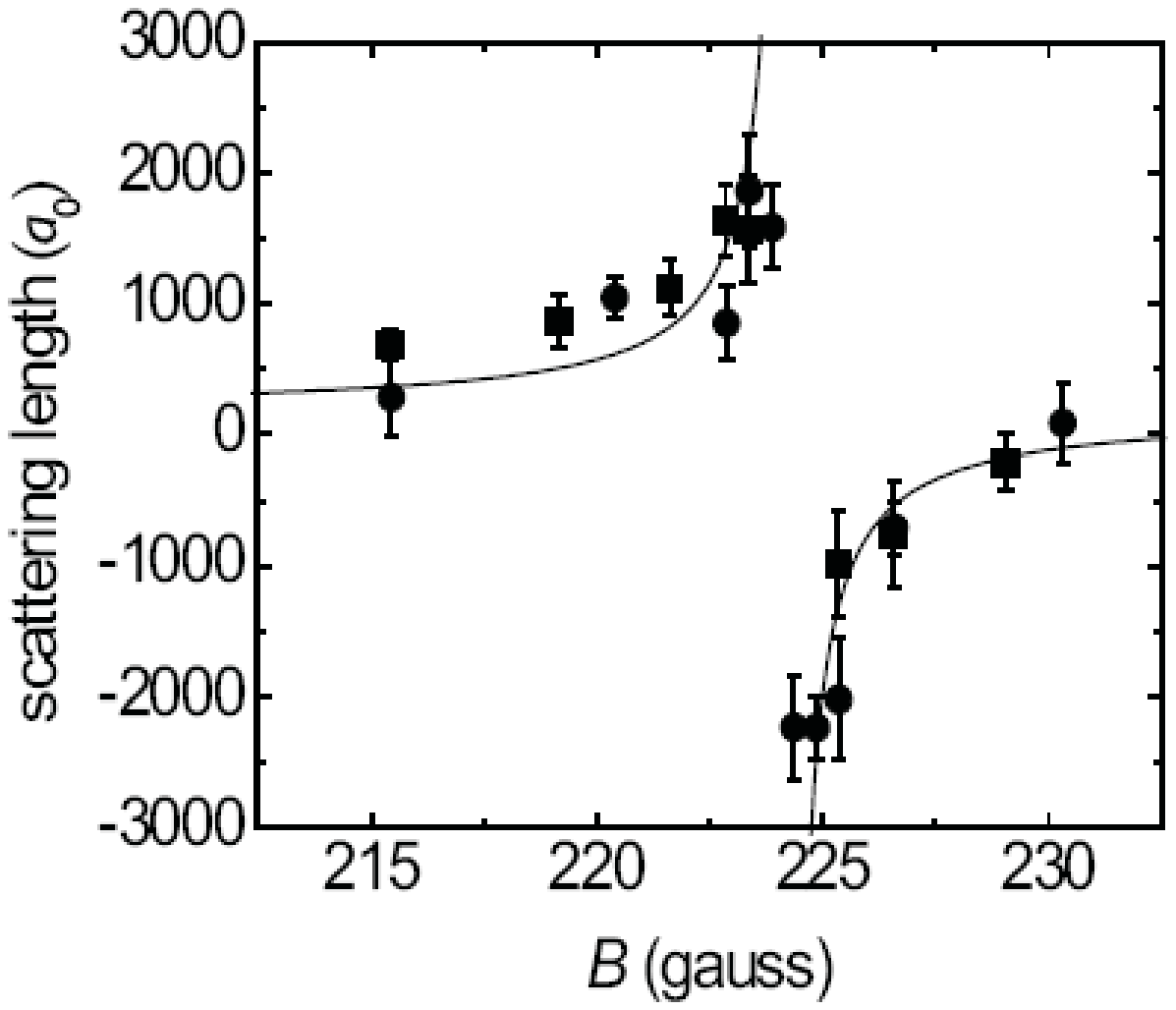}\ec
\end{minipage}\begin{minipage}{0.49\hsize}
\includegraphics[width=0.95\hsize]{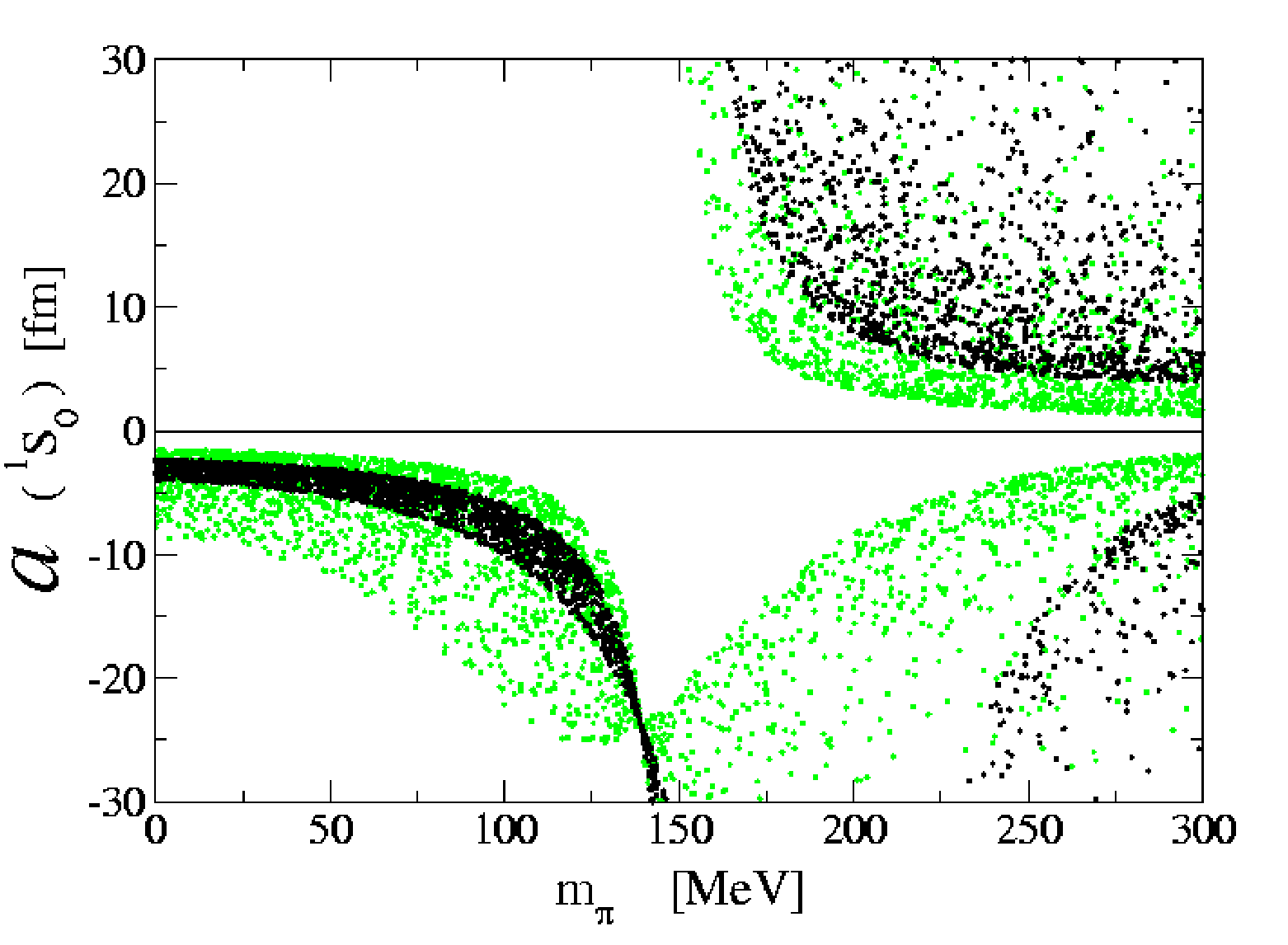}
\end{minipage}
\caption{\label{fig_fesh}
The left panel shows the scattering length of $^{40}$K Atoms
as a function of the magnetic field near a Feshbach resonance, from 
Regal (2005). The right panel shows the nucleon-nucleon scattering 
length in the $^1S_0$ channel as a function of the pion mass. The 
scatter plot indicates the uncertainty due to higher order terms
in the chiral effective lagrangian. Figure from Savage and Beane 
(2003).}
\end{figure}
%%%%%%%%%%%%%%%%%%%%%%%%%%%%%%%%%%%%%%%%%%%%%%%%%%%%%%%%%%%%%%%%%%%%

 A dilute gas of fermions in the unitarity limit is a strongly coupled
quantum liquid that exhibits many interesting properties. On a qualitative
level these properties are of interest in connection with other strongly 
interacting field theories. The unitarity limit is also quantitatively 
relevant to the physics of dilute neutron matter. The neutron-neutron
scattering length is $a_{nn}=-18$ fm and the effective range is $r_{nn}
=2.8$ fm. This means that there is a range of densities for which the 
interparticle spacing is large compared to the effective range but 
small compared to the scattering length. It is interesting to note that
the neutron scattering length depends on the quark masses in a way 
that is very similar to the dependence of atomic scattering lengths 
on the magnetic field near a Feshbach resonance \cite{Beane:2002xf}, 
see Fig.~\ref{fig_fesh}.

%%%%%%%%%%%%%%%%%%%%%%%%%%%%%%%%%%%%%%%%%%%%%%%%%%%%%%%%%%%%%%%%%%%%
\section{Universal equation of State}
\label{sec_eos}
\subsection{Cold Atomic Gases}
\label{sec_eos_unit}
%%%%%%%%%%%%%%%%%%%%%%%%%%%%%%%%%%%%%%%%%%%%%%%%%%%%%%%%%%%%%%%%%%%%

 We first consider the equation of state of cold fermions in the
unitarity limit. We are interested in the limit $(k_Fa)\to\infty$ 
and $(k_Fr)\to 0$, where $k_F$ is the Fermi momentum, $a$ is the 
scattering length and $r$ is the effective range. From dimensional 
analysis it is clear that the energy per particle has to be 
proportional to energy per particle of a free Fermi gas at the 
same density
\be
\frac{E}{A} = \xi \Big(\frac{E}{A}\Big)_0 = \xi 
\frac{3}{5}\Big(\frac{k_F^2}{2m}\Big),
\ee
where $k_F$ is the Fermi momentum. The calculation of the dimensionless 
quantity $\xi$ is a non-perturbative problem. We shall tackle this problem 
using a combination of effective field theory and lattice field theory 
methods. We first observe that in the low density limit the details of 
the interaction are not important. The physics of the unitarity limit 
is captured by an effective lagrangian of point-like fermions interacting 
via a short-range interaction. The lagrangian is 
\be 
\label{l_4f}
{\mathcal L} = \psi^\dagger \left( i\partial_0 +
 \frac{\nabla^2}{2m} \right) \psi 
 - \frac{C_0}{2} \left(\psi^\dagger \psi\right)^2 ,
\ee
where $m$ is the mass of the fermion and $C_0$ is the strength of the 
four-fermion interaction. In the weak coupling limit $C_0$ is related 
to the scattering length by $C_0=4\pi a/m$. 

%%%%%%%%%%%%%%%%%%%%%%%%%%%%%%%%%%%%%%%%%%%%%%%%%%%%%%%%%%%%%%%%%%%%
\begin{figure}[t]
\bc\includegraphics[angle=-90,width=0.75\hsize]{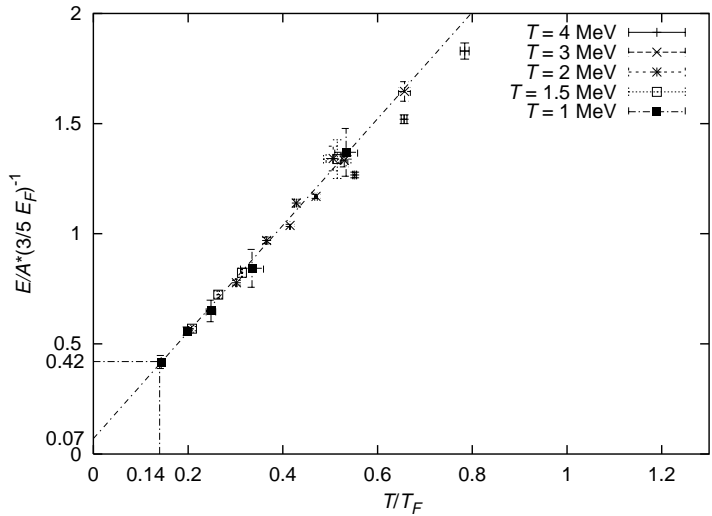}\ec
\caption{\label{fig_latt}
Lattice results for the energy per particle of a dilute Fermi gas 
from Lee \& Sch\"afer (2005). We show the energy per particle 
in units of $3E_F/5$ as a function of temperature in units of 
$T_F$.}
\end{figure}
%%%%%%%%%%%%%%%%%%%%%%%%%%%%%%%%%%%%%%%%%%%%%%%%%%%%%%%%%%%%%%%%%%%%

 If the scattering length is small then the energy of the many body 
system can be computed as a perturbative expansion in $(k_Fa)$. If 
$(k_Fa)$ is large we have to rely on numerical simulations. The 
partition function can be written as \cite{Lee:2004qd}
\be
Z = \int DsDcDc^{\ast}\exp\left[  -S\right]  ,
\ee
where $S$ is a discretized euclidean action
\bea
S  &=& 
 \sum_{\vec{n},i}\left[  e^{-\mu\alpha_{t}}c_{i}^{\ast}
   (\vec{n})c_{i}^{\prime}(\vec{n}+\hat{0})-e^{\sqrt{-C_0\alpha_{t}}
  s(\vec{n})+\frac{C_0\alpha_{t}}{2}}(1-6h)c_{i}^{\ast}
   (\vec{n})c_{i}^{\prime}(\vec{n})\right] \nonumber\\
& & \hspace{0.3cm}\mbox{} 
   -h\sum_{\vec{n},l_{s},i}\left[  
   c_{i}^{\ast}(\vec{n})c_{i}^{\prime}(\vec{n}
   +\hat{l}_{s})+c_{i}^{\ast}(\vec{n})c_{i}^{\prime}
  (\vec{n}-\hat{l}_{s})\right]  +\frac{1}{2}\sum_{\vec{n}}s^{2}(\vec{n}).
\eea
Here, $s$ is a Hubbard-Stratonovich field, $c$ is a Grassmann field, 
$\alpha_t$ is the ratio of the temporal and spatial lattice spacings and 
$h=\alpha_t/(2m)$. The sums are over spin labels $i$, lattice sites
$\vec{n}$, and spatial unit vectors $\hat{l}$. $\hat{0}$ is a temporal 
unit vector. Note that for $C_0<0$ the action is real and standard 
Monte Carlo simulations are possible. Results in the unitarity limit are 
shown in Fig.~\ref{fig_latt}. From these simulations we concluded that 
$\xi=(0.09-0.42)$. Lee performed canonical simulations at $T=0$ and 
obtained \cite{Lee:2005fk} $\xi=0.25$ . Green Function Monte Carlo 
calculations give \cite{Carlson:2003wm} $\xi=0.44$, and finite 
temperature lattice simulations have been extrapolated to $T=0$ to 
yield similar results \cite{Bulgac:2005pj,Burovski:2006}.

%%%%%%%%%%%%%%%%%%%%%%%%%%%%%%%%%%%%%%%%%%%%%%%%%%%%%%%%%%%%%%%%%%%%
\begin{figure}[t]
\bc\includegraphics[width=0.75\hsize]{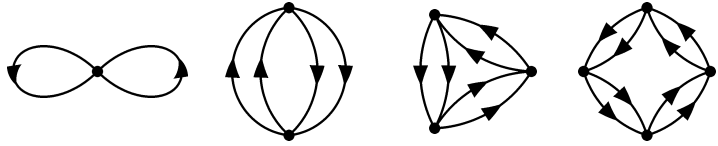}\ec
\caption{\label{fig_lad}
Particle-particle ladder diagrams.}
\end{figure}
%%%%%%%%%%%%%%%%%%%%%%%%%%%%%%%%%%%%%%%%%%%%%%%%%%%%%%%%%%%%%%%%%%%%

 It is also interesting to find analytical approaches to the equation 
of state in the unitarity limit. Since the two-body interaction is large 
it is natural to begin with the sum of all two-body diagrams, see 
Fig.~\ref{fig_lad}. This sum gives \cite{Schafer:2005kg}
\be
\label{pp_lad}
\frac{E}{N} =\frac{k_F^2}{2M}\left\{ \frac{3}{5} + 
 \frac{2(k_Fa)/(3\pi)}{1-\frac{6}{35\pi}(11-2\log(2))(k_Fa)}\right\}.
\ee
from which we deduce $\xi\simeq 0.32$. This is reasonably close to
the numerical results, but since the system is strongly correlated 
there is no obvious reason to restrict ourselves to two-body ladders. 
We have recently studied the possibility that equ.~(\ref{pp_lad}) 
can be justified as the leading term in an expansion in $1/d$, where 
$d$ is the number of spatial dimensions \cite{Steele:2000qt,Schafer:2005kg}.
This approach appears promising, but $1/d$ corrections have not been 
worked out yet. In order to obtain a smooth $d\to\infty$ limit the 
coupling constant has to be scaled in a specific way. Nussinov
\& Nussinov observed that if the limit $a\to\infty$ is taken first 
then the point $d=4$ is special \cite{Nussinov:2004}. In $d=4$ the 
two-body wave function at $a=\infty$ has a $1/r^2$ behavior and the 
normalization is concentrated near the origin. This implies that the 
many body system is equivalent to a gas of non-interacting bosons and 
$\xi(d\!=\!4)=0$. Nishida and Son computed $\xi$ in $d=3$ using an 
expansion around this limit. At next-to-leading order they find 
\cite{Nishida:2006br} $\xi=0.475$.

%%%%%%%%%%%%%%%%%%%%%%%%%%%%%%%%%%%%%%%%%%%%%%%%%%%%%%%%%%%%%%%%%%%%
\subsection{Strongly Coupled Gauge Theory}
\label{sec_susy}
%%%%%%%%%%%%%%%%%%%%%%%%%%%%%%%%%%%%%%%%%%%%%%%%%%%%%%%%%%%%%%%%%%%%

  In connection with the RHIC program we are interested in understanding
the quark gluon plasma in the vicinity of the critical temperature. 
From lattice simulations we know that the energy density quickly 
reaches about 80\% of the ideal gas value, and that this ratio is 
only very weakly temperature dependent on temperature for $T>2T_c$. 
This behavior can be understood in hard thermal loop resummed perturbation 
theory \cite{Blaizot:2003tw}. In this framework the degrees of freedom
are dressed quasi-quarks and quasi-gluons, and these quasi-particles
are weakly interacting.

 Transport properties of the plasma indicate that this may not be the 
end of the story. If quasi-particles are indeed weakly interacting then 
it is hard to see how the viscosity can be anomalously low or the 
opacity be very large. A complementary, strong coupling, calculation was 
performed in the strong coupling and large $N_c$ limit of ${\cal N}=4$ 
SUSY Yang Mills theory. The calculation is based on the duality
between the strongly coupled gauge theory and weakly coupled 
string theory on $AdS_5\times S_5$ discovered by Maldacena
\cite{Maldacena:1997re}. The correspondence can be extended to 
finite temperature. In this case the relevant configurations is
an $AdS_5$ black hole. The temperature of the gauge theory is 
given by the Hawking temperature of the black hole, and the 
entropy is given by the Hawking-Beckenstein formula. The result
is that the entropy density of the strongly coupled field theory
is equal to 3/4 of the free field theory value \cite{Gubser:1998nz}.
Clearly, the number 3/4 can be viewed as a gauge theory analog 
of the parameter $\xi$ studied in the previous section. The 
remarkable result is that the gauge theory value is so close 
to one, so that based on the equation of state alone one cannot 
easily distinguish a strongly and a weakly coupled system. 

%%%%%%%%%%%%%%%%%%%%%%%%%%%%%%%%%%%%%%%%%%%%%%%%%%%%%%%%%%%%%%%%%%%%
\section{Pairing}
\label{sec_pair}
\subsection{Cold Atomic Gases}
\label{sec_pair_unit}
%%%%%%%%%%%%%%%%%%%%%%%%%%%%%%%%%%%%%%%%%%%%%%%%%%%%%%%%%%%%%%%%%%%%

 At low temperature the atomic gas becomes superfluid. If the 
coupling is weak then the gap and the critical temperature can be 
calculated using BCS theory. The result is 
\be
\label{gap_bcs}
\Delta = \frac{8E_F}{(4e)^{1/3}e^2}
   \exp\left(-\frac{\pi}{2k_F|a|}\right),
\ee
where the factor $(4e)^{1/3}$ is the screening correction first 
computed by Gorkov et al.~\cite{Gorkov:1961}. Higher order corrections 
are suppressed by powers of $(k_Fa)$. In BCS theory the critical 
temperature is given by $T_c=e^\gamma\Delta/\pi$. Clearly, the 
critical temperature grows with the scattering length. Naively 
extrapolating equ.~(\ref{gap_bcs}) to the unitarity limit gives 
$T_c\simeq 0.28 E_F$. In the BEC limit the critical temperature 
is given by the Einstein result $T_c=3.31\rho_M^{2/3}/m_M$, where 
$\rho_M=\rho/2$ and $m_M=2m$ are the density and mass of the molecules. 
This implies that $T_c\simeq 0.21 E_F$ where we have defined the 
Fermi energy as $E_F=k_F^2/(2m)=(3\pi^2\rho)^{2/3}/(2m)$. Interactions 
between the bosons increase the critical temperature \cite{Baym:2001}, 
$T_c=T_c^0+O(a_B\rho^{1/3})$. Near the unitarity limit the boson 
scattering length $a_B$ is proportional to the scattering length
between the fermions \cite{Petrov:2003}, $a_{BB}\simeq 0.6a$. These 
results suggest that the unitarity limit corresponds to the maximum 
possible value of $T_c/E_F$.

 The value of $T_c$ has been determined in a number of lattice 
calculations. A careful scaling analysis by Burovski et al.~yields
\cite{Burovski:2006} $T_c=0.152(7)E_F$. Earlier determinations of 
$T_c$ can be found in \cite{Wingate:2004wm,Bulgac:2005pj}. A direct 
calculation of the pairing gap at zero temperature using Green Function 
Monte Carlo methods gives \cite{Chang:2004sj} $\Delta=0.54E_F$. 

%%%%%%%%%%%%%%%%%%%%%%%%%%%%%%%%%%%%%%%%%%%%%%%%%%%%%%%%%%%%%%%%%%%%
\subsection{Color Superconductivity}
\label{sec_csc}
%%%%%%%%%%%%%%%%%%%%%%%%%%%%%%%%%%%%%%%%%%%%%%%%%%%%%%%%%%%%%%%%%%%%

 Pairing also takes place in the high density, low temperature 
phase of QCD. At asymptotically large density the attraction 
is due to one-gluon exchange between fermions with opposite momenta
and anti-symmetric spin, color and flavor wave functions. In this 
limit the pairing gap is given by 
\cite{Son:1999uk,Schafer:1999jg}
\be
\label{gap_oge}
\Delta = 2\Lambda_{BCS}
   \exp\left(-\frac{\pi^2+4}{8}\right)
   \exp\left(-\frac{3\pi^2}{\sqrt{2}g}\right).
\ee
where $g$ is the running coupling constant evaluated at the scale 
$\mu$ and $\Lambda_{BCS}=256\pi^4(2/N_f)^{5/2}g^{-5}\mu$. Here,
$\mu$ is the baryon chemical potential and $N_F$ is the number 
of flavors. This result exhibits a non-BCS like dependence on the 
coupling constant which is related to the presence of unscreened
magnetic gluon exchanges. The critical temperature is nevertheless
given by the BCS result $T_c=e^\gamma\Delta/\pi$. 

 If the weak coupling limit the gap and the critical temperature are 
exponentially small. The ratio $T_c/E_F$ increases with $g$ and reaches 
a maximum of $T_c=0.025E_F$ at $g=4.2$. The maximum occurs at strong 
coupling and the result is not reliable. Using phenomenological 
interactions, or extrapolating the QCD Dyson-Schwinger equations 
into the strong coupling domain \cite{Nickel:2006vf}, one finds 
critical temperatures as large as $T_c=0.15E_F$. In connection
with phenomenological applications to neutron stars and the physics
of heavy ion collisions in the regime of the highest baryon densities
it is very important to reduce the uncertainty in these estimates. 
One possibility is to use lattice studies of QCD-like theories in which 
the euclidean action remains positive at finite chemical potential. 
Examples are QCD with two colors and QCD at finite isospin density. 

%%%%%%%%%%%%%%%%%%%%%%%%%%%%%%%%%%%%%%%%%%%%%%%%%%%%%%%%%%%%%%%%%%%%
\section{Stressed Pairing}
\label{sec_stress}
%%%%%%%%%%%%%%%%%%%%%%%%%%%%%%%%%%%%%%%%%%%%%%%%%%%%%%%%%%%%%%%%%%%%

%%%%%%%%%%%%%%%%%%%%%%%%%%%%%%%%%%%%%%%%%%%%%%%%%%%%%%%%%%%%%%%%%%%%
\subsection{Polarized Cold Atomic Gases}
\label{sec_stress_unit}
%%%%%%%%%%%%%%%%%%%%%%%%%%%%%%%%%%%%%%%%%%%%%%%%%%%%%%%%%%%%%%%%%%%%

\begin{figure}[t]
\bc\includegraphics[width=7.5cm]{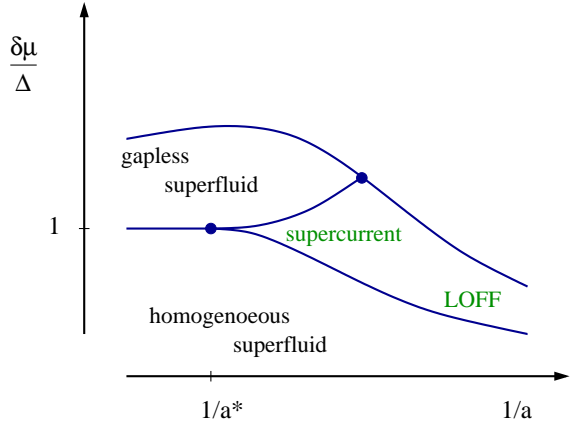}\ec
\caption{\label{fig_ph}
Conjectured phase diagram for a polarized cold atomic Fermi gas
as a function of the scattering length $a$ and the difference 
in the chemical potentials $\delta\mu=\mu_\uparrow-\mu_\downarrow$, 
from Son \& Stephanov (2005).}
\end{figure}

 The superfluid state discussed in Sect.~\ref{sec_pair_unit} involves
equal numbers of spin up and spin down fermions. One aspect of the 
paired state that has attracted a lot of interest is the response 
of this system to a non-zero chemical potential coupled to the 
third component of spin, $\delta\mu=\mu_\uparrow-\mu_\downarrow$.
A conjectured (and, most likely, oversimplified) phase diagram 
for a polarized gas is shown in Fig.~\ref{fig_ph}. In the BEC 
limit the gas consists of tightly bound spin singlet molecules. 
Adding an extra up or down spin requires energy $\Delta$. For 
$|\delta\mu|>\Delta$ the system is a homogeneous mixture of a Bose 
condensate and a fully polarized Fermi gas. One can show that in the 
dilute limit this mixture is stable with regard to phase separation 
\cite{Viverit:2000}.

%%%%%%%%%%%%%%%%%%%%%%%%%%%%%%%%%%%%%%%%%%%%%%%%%%%%%%%%%%%%%%%%%%%%
%%%%%%%%%%%%%%%%%%%%%%%%%%%%%%%%%%%%%%%%%%%%%%%%%%%%%%%%%%%%%%%%%%%%

 We can also analyze the system in the BCS limit. This analysis
goes back to the work of Larkin, Ovchninikov, Fulde and Ferell (LOFF) 
\cite{Larkin:1964,Fulde:1964}, see the review \cite{Casalbuoni:2003wh}.   
We first consider homogeneous solutions to the BCS gap equation for 
$\delta\mu\neq 0$. In the regime $\delta\mu<\Delta_0$ where $\Delta_0=
\Delta(\delta\mu\!=\!0)$ the gap equation has a solution with gap 
parameter $\Delta=\Delta_0$. This solution is stable if $\delta\mu<
\Delta_0/\sqrt{2}$ but only meta-stable in the regime $\Delta_0/\sqrt{2}
<\delta\mu<\Delta_0$. The BCS solution has vanishing polarization. 
The transition to a polarized normal phase is first order, and 
systems at intermediate polarization correspond to mixed phases.

LOFF~studied whether it is possible to find a stable solution in 
which the gap has a spatially varying phase
\be 
\label{loff}
 \Delta(\vec{x})= \Delta e^{2i\vec{q}\cdot\vec{x}}.
\ee
This solution exists in the LOFF window $\delta\mu_1<\delta\mu<
\delta\mu_2$ with $\delta\mu_1=\Delta_0/\sqrt{2}\simeq 0.71
\Delta_0$ and $\delta\mu_2\simeq 0.754\Delta_0$. The LOFF momentum 
$q$ depends on $\delta\mu$. Near $\delta\mu_2$ we have $qv_F
\simeq 1.2\delta\mu$, where $v_F$ is the Fermi velocity. The gap 
$\Delta$ goes to zero near $\delta\mu_2$ and reaches $\Delta\simeq 
0.25\Delta_0$ at $\delta\mu_1$.

  These results suggest that at some point on the phase diagram between 
the BEC and BCS limits the homogeneous superfluid becomes unstable with 
respect to the formation of a non-zero supercurrent $\vec\nabla\varphi$,
where $\varphi$ is the phase of the condensate. We can study this 
question using the effective lagrangian
\be
\label{leff_gbcs}
  {\cal L} =  \psi^\dagger \Big(i\partial_0 - \epsilon(-i\vec{\partial}) 
  - i(\vec{\partial}\varphi)\cdot
   \frac{\stackrel{\scriptstyle\leftrightarrow}{\partial}}{2m} 
    \Big)\psi
  + \frac{f_t^2}{2} \dot\varphi^2 - \frac{f^2}{2} (\vec{\partial}\varphi)^2.
\ee
Here, $\psi$ describes a fermion with dispersion law $\epsilon(\vec{p})$ 
and $\varphi$ is the superfluid Goldstone mode. The low energy 
parameters $f_t$ and $f$ are related to the density and the velocity of 
sound. The p-wave coupling of the fermions to the Goldstone boson is 
governed by the $U(1)$ symmetry of the theory. 

 Setting up a current $\vec{v}_s=\vec{\partial}\varphi/m$ requires energy
$f^2m^2v_s^2/2$. The contribution from fermions can be computed using 
the fermion dispersion law in the presence of a non-zero current
\be
  \epsilon_v(\vec{p}) = \epsilon(\vec{p})
          + \vec{v}_s\cdot\vec{p}  -\delta\mu\, .
\ee
The total free energy is 
\be 
F(v_s) = \frac{1}{2} n m v_s^2 + 
  \int \frac{d^3p}{(2\pi)^3} \, \epsilon_v(\vec{p}) 
    \Theta\left(-\epsilon_v(\vec{p})\right),
\ee
where $n$ is the density and we have used $f^2=n/m$. Son and Stephanov 
noticed that the stability of the homogeneous phase depends crucially on 
the nature of the dispersion law $\epsilon(p)$. For small momenta we 
can write $\epsilon(p)\simeq \epsilon_0+\alpha p^2+\beta p^4$. In the 
BEC limit $\alpha>0$ and the minimum of the dispersion curve is at 
$p=0$ while in the BCS limit $\alpha<0$ and the minimum is at $p\neq 0$.
In the latter case the density of states on the Fermi surface is finite 
and the system is unstable with respect to the formation of a non-zero 
current. On the other hand, if the minimum of the dispersion curve is 
at zero, then the density of states vanishes and there is no instability. 
As a consequence there is a critical point along the BEC-BCS 
line at which the instability will set in. 
 
 Clearly, the LOFF solution is of the same type as the supercurrent 
state. The difference is that in the supercurrent state the current 
is much smaller than the gap, $v_F(\nabla\varphi)\ll \Delta$, while 
in the LOFF phase $v_F(\nabla\varphi)> \Delta$ (and $v_F(\nabla\varphi)
\gg \Delta$ near $\delta\mu_2$). In the supercurrent state the Fermi 
surface is mostly gapped but a small shell near one of the pole caps is
ungapped. In the weakly coupled LOFF state there are gapless excitations 
over most of the Fermi surface but pairing takes place near two rings 
on the northern and southern hemisphere. 

%%%%%%%%%%%%%%%%%%%%%%%%%%%%%%%%%%%%%%%%%%%%%%%%%%%%%%%%%%%%%%%%%%%%
\subsection{CFL Phase at Non-zero Strange Quark Mass}
\label{sec_stress_cfl}
%%%%%%%%%%%%%%%%%%%%%%%%%%%%%%%%%%%%%%%%%%%%%%%%%%%%%%%%%%%%%%%%%%%%

 The exact nature of the color superconducting phase in QCD depends
on the baryon chemical potential, the number of quark flavors and 
on their masses. If the baryon chemical is much larger than the 
quark masses then the ground state of QCD with three flavors is the 
color-flavor-locked (CFL) phase. The CFL phase is characterized by 
the pair condensate \cite{Alford:1999mk}
\be
\label{cfl}
 \langle \psi^a_i C\gamma_5 \psi^b_j\rangle  =
  (\delta^a_i\delta^b_j-\delta^a_j\delta^b_i) \phi .
\ee
This condensate leads to a gap in the excitation spectrum
of all fermions and completely screens the gluonic interaction.
Both the chiral $SU(3)_L\times SU(3)_R$ and color $SU(3)$
symmetry are broken, but a vector-like $SU(3)$ flavor symmetry
remains unbroken.

In the real world the strange quark mass is not equal to the masses
of the up and down quark and flavor symmetry is broken. At high baryon
density the effect of the quark masses is governed by the shift
$\mu_q = m_q^2/(2\mu)$ of the Fermi energy due to the quark mass.
This implies that flavor symmetry breaking becomes more important as
the density is lowered. Clearly, the most important effect is due
to the strange quark mass. There are two scales that are important
in relation to $\mu_s$. The first is the mass of the lightest
strange Goldstone boson, the kaon, and the second is the gap $\Delta$
for fermionic excitations. When $\mu_s$ becomes equal to $m_K$ the 
CFL phase undergoes a transition to a phase with kaon condensation
\cite{Bedaque:2001je}. In the following we shall study the phase
structure near $\mu_s\sim\Delta$.

  Our starting point is the effective theory of the CFL phase derived 
in \cite{Casalbuoni:1999wu,Kryjevski:2004jw}. The effective lagrangian 
contains Goldstone boson fields $\Sigma$ and baryon fields $N$. The meson 
fields arise from chiral symmetry breaking in the CFL phase. The leading 
terms in the effective theory are
\be
\label{l_mes}
{\cal L} =  \frac{f_\pi^2}{4} {\rm Tr}\left(
  \nabla_0 \Sigma \nabla_0 \Sigma^{\dagger} - v_\pi^2
  \vec{\nabla} \Sigma \vec{\nabla}\Sigma^\dagger \right),
\ee
where $f_\pi$ is the pion decay constant. The chiral field transforms as 
$\Sigma\to L\Sigma R^\dagger$ under chiral transformations $(L,R)\in 
SU(3)_L\times SU(3)_R$. Baryon fields originate from quark-hadron 
complementarity \cite{Schafer:1998ef}. The effective lagrangian is 
\bea
\label{l_bar}
 {\cal L} =
 {\rm Tr}\left(N^\dagger iv^\mu D_\mu N\right)
 - D{\rm Tr} \left(N^\dagger v^\mu\gamma_5
               \left\{ {\cal A}_\mu,N\right\}\right)
 - F{\rm Tr} \left(N^\dagger v^\mu\gamma_5
               \left[ {\cal A}_\mu,N\right]\right)
  \nonumber \\
   \mbox{} + \frac{\Delta}{2} \left\{
     \left( {\rm Tr}\left(N_LN_L \right)
   - \left[ {\rm Tr}\left(N_L\right)\right]^2 \right)
   - \left( {\rm Tr} \left(N_RN_R \right)
   - \left[ {\rm Tr}\left(N_R\right)\right]^2 \right)
     + h. c.  \right\},
\eea
where $N_{L,R}$ are left and right handed baryon fields in the adjoint 
representation of flavor $SU(3)$, $v^\mu=(1,\vec{v})$ is the Fermi 
velocity, and $\Delta$ is the superfluid gap. We can think of $N$
as being composed of a quark and a diquark field, $N_L\sim q_L
\langle q_Lq_L\rangle$. The interaction of the baryon field with 
the Goldstone bosons is dictated by chiral symmetry. The covariant 
derivative is given by $D_\mu N=\partial_\mu N +i[{\cal V}_\mu,N]$. 
The vector and axial-vector currents are
\be
 {\cal V}_\mu = -\frac{i}{2}\left\{
  \xi \partial_\mu\xi^\dagger +  \xi^\dagger \partial_\mu \xi
  \right\}, \hspace{1cm}
{\cal A}_\mu = -\frac{i}{2} \xi\left(\partial_\mu
    \Sigma^\dagger\right) \xi ,
\ee
where $\xi$ is defined by $\xi^2=\Sigma$. The low energy constants 
$f_\pi,v_\pi,D,F$ can be calculated in perturbative QCD. Symmetry 
arguments can be used to determine the leading mass terms in the 
effective lagrangian. Bedaque and Sch\"afer observed that $X_L=
MM^\dagger/(2p_F)$ and $X_R=M^\dagger M/(2p_F)$ act as effective 
chemical potentials and enter the theory like the temporal components 
of left and right handed flavor gauge fields \cite{Bedaque:2001je}. 
We can make the effective lagrangian invariant under this symmetry 
by introducing the covariant derivatives
\bea
\label{V_X}
 D_0N &=& \partial_0 N+i[\Gamma_0,N], \hspace{0.5cm}
 \Gamma_0 = -\frac{i}{2}\left\{
  \xi \left(\partial_0+ iX_R\right)\xi^\dagger +
  \xi^\dagger \left(\partial_0+iX_L\right) \xi
  \right\}, \\
\nabla_0\Sigma &=& \partial_0\Sigma+iX_L\Sigma-i\Sigma X_R.
\eea
Using equ.~(\ref{l_bar}-\ref{V_X}) we can calculate the dependence
of the gap in the fermion spectrum on the strange quark mass. For
$m_s=0$ there are 8 quasi-particles with gap $\Delta$ and one
quasi-particle with gap $2\Delta$. As $m_s$ increases some of 
the gaps decrease. In the $K^0$ condensed phase the gap of the 
lowest mode is approximately given by $\Delta=\Delta_0-3\mu_s/4$ 
where $\mu_s=m_s^2/(2p_F)$ and $\Delta_0$ is the gap in the chiral 
limit. 

%%%%%%%%%%%%%%%%%%%%%%%%%%%%%%%%%%%%%%%%%%%%%%%%%%%%%%%%%%%
\begin{figure}[t]
\bc\includegraphics[width=6.3cm]{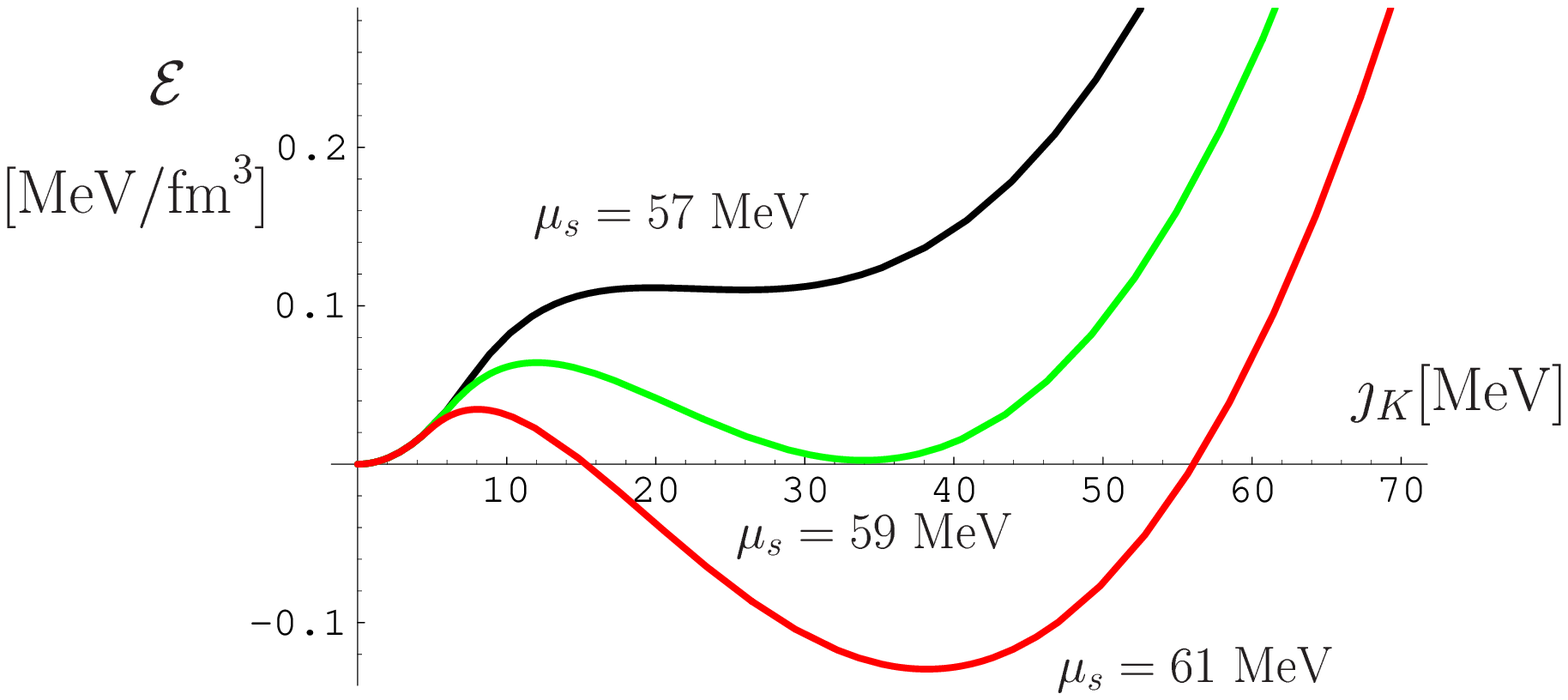}\hspace*{-0.3cm}
\includegraphics[width=6.3cm]{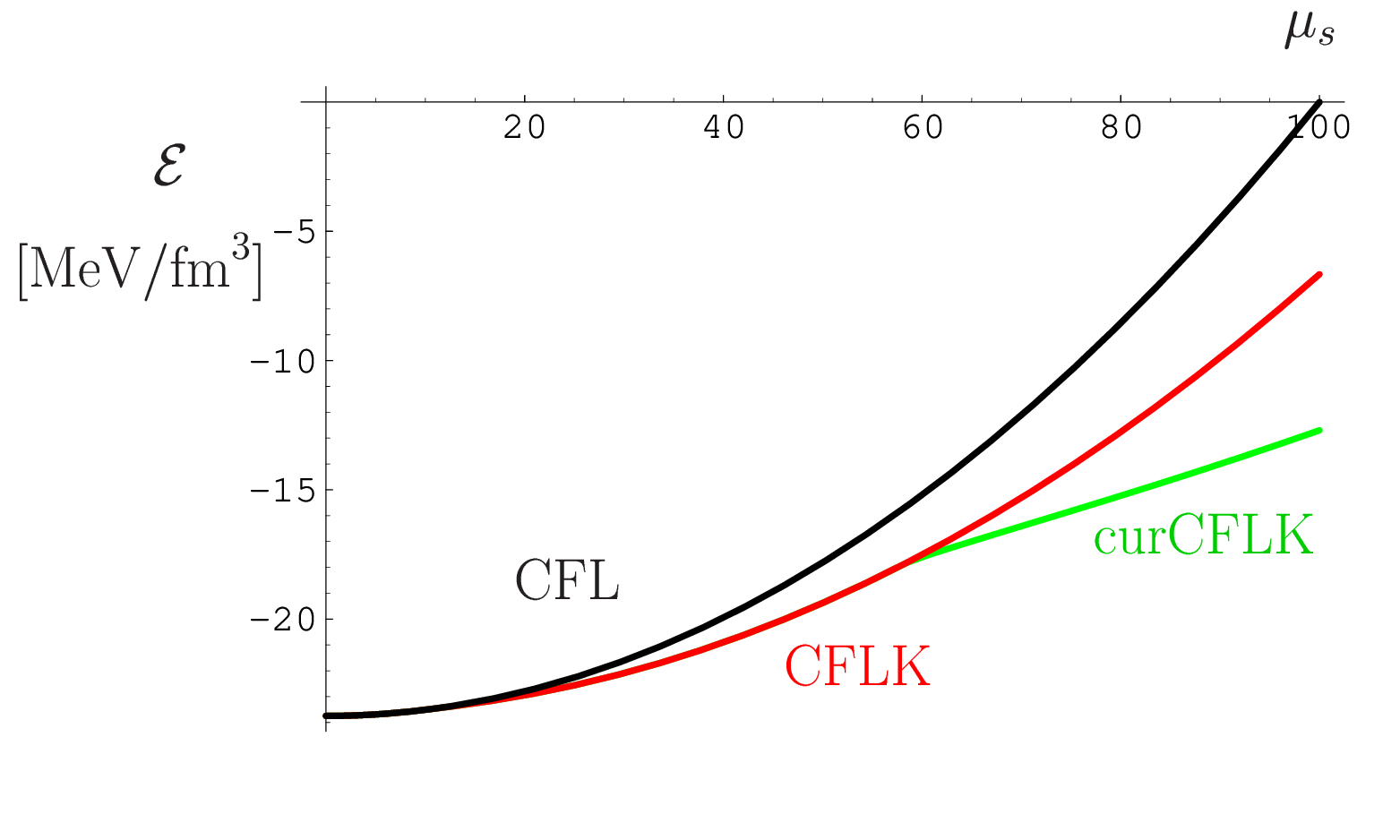}\ec
\caption{Left panel: Energy density as a function of the current 
$\jmath_K$ for several different values of $\mu_s=m_s^2/(2p_F)$ 
close to the phase transition. Right panel: Ground state energy 
density as a function of $\mu_s$. We show the CFL phase, the 
kaon condensed CFL (KCFL) phase, and the supercurrent state
(curKCFL).}
\label{fig_jfct}
\end{figure}
%%%%%%%%%%%%%%%%%%%%%%%%%%%%%%%%%%%%%%%%%%%%%%%%%%%%%%%%%%%

 For $\mu_s\sim 4\Delta_0/3$ the system contains low energy
fermions derivatively coupled to Goldstone bosons. The situation 
is essentially equivalent to the cold atomic system studied in the 
previous section. The main difference is that in the CFL phase
there are many more currents that can appear. In the $K^0$ 
condensed phase the natural ansatz is a hypercharge current carried 
by the kaon field. We take $\xi(x)= U(x)\xi_{K^0}U^\dagger (x)$ where 
$\xi_{K^0}$ is the $K^0$ background, $U(x)$ is a hypercharge 
transformation and $\vec\jmath_{K}=iU^\dagger\vec{\nabla}U$ is the 
kaon current. The dispersion relation of the lowest mode is 
\cite{Kryjevski:2005qq,Schafer:2005ym}
\be
\label{disp_ax}
\omega_l = \Delta_0 +\frac{l^2}{2\Delta_0}-\frac{3}{4}
  \mu_s -\frac{1}{4}\vec{v}\cdot\vec{\jmath}_K,
\ee
where $l$ is the momentum relative to the Fermi surface. The energy 
relative to the CFL phase is the kinetic energy of the current plus 
the energy of occupied gapless modes
\be
\label{efct}
{\cal E} = \frac{1}{2}v_\pi^2f_\pi^2\vec{\jmath}_K^{\, 2} 
 + \frac{\mu^2}{\pi^2}\int dl \int
 \frac{d\Omega}{4\pi} \;\omega_l \theta(-\omega_l) .
\ee
The energy functional can develop a minimum at non-zero $\jmath_K$ 
because the current lowers the energy of the fermions near one of 
the pole caps on the Fermi surface. Introducing the dimensionless 
variables $x=\jmath_K/(a\Delta)$ and $h=(3\mu_s-4\Delta)/(a\Delta)$
we can write
\be
\label{efct_x}
{\cal E} = Cf_h(x), \hspace{0.5cm}
 f_h(x) = x^2-\frac{1}{x}\left[
   (h+x)^{5/2}\Theta(h+x) - (h-x)^{5/2}\Theta(h-x) \right] ,
\ee
where $C$ and $a$ are numerical constants. The functional 
given in equ.~(\ref{efct_x}) was analyzed in 
\cite{Son:2005qx,Kryjevski:2005qq,Schafer:2005ym}, see Fig.~\ref{fig_jfct}. 
There is a critical chemical potential $\mu_s= (4/3+ah_{crit}/3)\Delta$ 
above which the groundstate contains a non-zero supercurrent $\jmath_K$. 
This current is canceled by a backflow of gapless fermions. The Goldstone
current phase is analogous to the supercurrent state in cold atomic systems, 
and to $p$-wave pion condensates in nuclear matter. 

 The Goldstone current phase in pure CFL matter, without an $s$-wave kaon 
condensate, was analyzed by Gerhold and Sch\"afer \cite{Gerhold:2006dt}.
In that case the structure of the current is different, but the 
energy functional and the nature of the instability are very similar. 
Gerhold and Sch\"afer showed, in particular, that the magnetic
screening masses in the Goldstone boson current phase are real. 
Near the onset of the instability the current is small compared 
to the gap, but the current grows with the mismatch $\mu_s$. Once 
the current becomes comparable to the gap it may be more appropriate 
to characterize the system as a LOFF state~\cite{Casalbuoni:2005zp}.

%%%%%%%%%%%%%%%%%%%%%%%%%%%%%%%%%%%%%%%%%%%%%%%%%%%%%%%%%%%%%%%%%%%%
\section{Outlook}
\label{sec_sum}
%%%%%%%%%%%%%%%%%%%%%%%%%%%%%%%%%%%%%%%%%%%%%%%%%%%%%%%%%%%%%%%%%%%%
 
 There are many questions about cold fermion gases in the unitarity
limit, and about their relevance to strongly interacting hadronic 
matter, that remain to be resolved. The unitarity limit is most 
directly connected with the physics of dilute neutron matter. 
In neutron matter the scattering length is not exactly infinite, 
and the effective range is not zero. It is clearly important to 
understand how corrections due to the finite scattering length 
and effective range affect the equation of state and other 
observables.

 There is a great deal of experimental effort devoted to the 
study of polarized fermionic gases. Mixed phases of superfluids
and polarized normal fluids have been observed \cite{Ketterle},
but so far none of the predicted inhomogeneous or gapless superfluids 
have been experimentally detected. Finally, much work remains to done 
with regard to the transport properties of cold fermionic gases. 
We would like to understand, both theoretically and experimentally,  
how the shear viscosity depends on the temperature and the scattering 
length. Damping of collective oscillations of trapped fermions
has been observed experimentally \cite{Thomas}, but the
damping coefficients have not been related to transport 
properties of the bulk system. 

Acknowledgments: The work described in this contribution was 
done in collaboration with S.~Cotanch, A.~Gerhold, C.-W.~Kao, 
A.~Kryjevski and D.~Lee. I would like to thank N.~Nygaard
for alerting me to an error in an earlier draft. This work 
is supported in part by the US Department of Energy grant 
DE-FG-88ER40388.


\begin{thebibliography}{0}
\bibitem{rhic:2005}
I.~Arsenne et al.~[Brahms],
B.~Back et al.~[Phobos],
K.~Adcox et al.~[Phenix],
J.~Adams et al.~[Star],
``First Three Years of Operation of RHIC'',
Nucl.\ Phys.\ {\bf A757}, 1-183 (2005).

\bibitem{Friedman:1981qw}
B.~Friedman and V.~R.~Pandharipande,
%``Hot And Cold, Nuclear And Neutron Matter,''
Nucl.\ Phys.\ A {\bf 361}, 502 (1981);
%%CITATION = NUPHA,A361,502;%%
S.~Y.~Chang {\it et al.},
%``Neutron Matter: A Superfluid Gas,''
Nucl.\ Phys.\ A {\bf 746}, 215 (2004)
[nucl-th/0401016];
%%CITATION = NUCL-TH 0401016;%%
G.~A.~Baker,
%``Neutron matter model,''
Phys.\ Rev.\ C {\bf 60}, 054311 (1999).
%%CITATION = PHRVA,C60,054311;%%

\bibitem{Rapp:1999qa}
R.~Rapp, T.~Sch\"afer, E.~V.~Shuryak and M.~Velkovsky,
%``High-density QCD and instantons,''
Annals Phys.\  {\bf 280}, 35 (2000)
[hep-ph/9904353];
%%CITATION = HEP-PH 9904353;%%
%\bibitem{Alford:1997zt}
M.~G.~Alford, K.~Rajagopal and F.~Wilczek,
%``QCD at finite baryon density: Nucleon droplets and color
%superconductivity,''
Phys.\ Lett.\ B {\bf 422}, 247 (1998)
[hep-ph/9711395].
%%CITATION = HEP-PH 9711395;%%

\bibitem{Regal:2005}
C.~Regal, Ph.~D.~Thesis, University of Colorado (2005), 
cond-mat/0601054.

\bibitem{Beane:2002xf}
S.~R.~Beane and M.~J.~Savage,
%``The quark mass dependence of two-nucleon systems,''
Nucl.\ Phys.\ A {\bf 717}, 91 (2003)
[nucl-th/0208021].
%%CITATION = NUCL-TH 0208021;%%

\bibitem{Lee:2004qd}
D.~Lee and T.~Sch{\"a}fer,
%``Neutron matter on the lattice with pionless effective field theory,''
Phys.\ Rev.\ C {\bf 72}, 024006 (2005)
[nucl-th/0412002];
%%CITATION = NUCL-TH 0412002;%%
%``Cold dilute neutron matter on the lattice I: Lattice virial coefficients
%and large scattering lengths,''
Phys.\ Rev.\ C {\bf 73}, 015201 (2006)
[nucl-th/0509017];
%%CITATION = NUCL-TH 0509017;%%
%``Cold dilute neutron matter on the lattice II: Results in the unitary
%limit,''
Phys.\ Rev.\ C {\bf 73}, 015202 (2006)
[nucl-th/0509018].
%%CITATION = NUCL-TH 0509018;%%

\bibitem{Lee:2005fk}
D.~Lee,
%``Ground state energy of spin-1/2 fermions in the unitary limit,''
Phys.\ Rev.\ B {\bf 73}, 115112 (2006)
[cond-mat/0511332].
%%CITATION = COND-MAT 0511332;%%

\bibitem{Carlson:2003wm}
J.~Carlson, J.~J.~Morales, V.~R.~Pandharipande and D.~G.~Ravenhall,
%``Quantum Monte Carlo Calculations of Neutron Matter,''
Phys.\ Rev.\ C {\bf 68}, 025802 (2003)
[nucl-th/0302041].
%%CITATION = NUCL-TH 0302041;%%

\bibitem{Bulgac:2005pj}
A.~Bulgac, J.~E.~Drut and P.~Magierski,
%``Spin 1/2 Fermions on a 3D-Lattice in the Unitary Regime at Finite
%Temperatures,''
Phys.\ Rev.\ Lett.\ {\bf 96}, 090404 (2006)
[cond-mat/0505374].
%%CITATION = COND-MAT 0505374;%%

\bibitem{Burovski:2006}
E.~Burovski, N.~Prokof'ev, B.~Svistunov, M.~Troyer,
%Critical Temperature and Thermodynamics of Attractive Fermions at Unitarity
Phys.\ Rev.\ Lett.\ {\bf 96}, 160402 (2006)
[cond-mat/0602224].
%%CITATION = COND-MAT 0602224;%%

\bibitem{Schafer:2005kg}
T.~Sch\"afer, C.~W.~Kao and S.~R.~Cotanch,
%``Many Body Methods and Effective Field Theory,''
Nucl.\ Phys.\ A {\bf 762}, 82 (2005)
[nucl-th/0504088].
%%CITATION = NUCL-TH 0504088;%%

\bibitem{Steele:2000qt}
J.~V.~Steele,
%``Effective Field Theory Power Counting at Finite Density,''
preprint, nucl-th/0010066.
%%CITATION = NUCL-TH 0010066;%%

\bibitem{Nussinov:2004}
Z.~Nussinov and S.~Nussinov,
preprint, cond-mat/0410597.

\bibitem{Nishida:2006br}
Y.~Nishida and D.~T.~Son,
%``An epsilon expansion for Fermi gas at infinite scattering length,''
preprint, cond-mat/0604500.
%%CITATION = COND-MAT 0604500;%%

\bibitem{Blaizot:2003tw}
J.~P.~Blaizot, E.~Iancu and A.~Rebhan,
Thermodynamics of the high-temperature quark gluon plasma,
in Quark Gluon Plasma 3, R.~Hwa, X.-N.~Wang, eds., (2003)
[hep-ph/0303185].
%%CITATION = HEP-PH 0303185;%%

\bibitem{Maldacena:1997re}
J.~M.~Maldacena,
%``The large N limit of superconformal field theories and supergravity,''
Adv.\ Theor.\ Math.\ Phys.\  {\bf 2}, 231 (1998)
[hep-th/9711200].
%%CITATION = HEP-TH 9711200;%%

\bibitem{Gubser:1998nz}
S.~S.~Gubser, I.~R.~Klebanov and A.~A.~Tseytlin,
%``Coupling constant dependence in the thermodynamics of N = 4  supersymmetric
%Yang-Mills theory,''
Nucl.\ Phys.\ B {\bf 534}, 202 (1998)
[hep-th/9805156].
%%CITATION = HEP-TH 9805156;%%

\bibitem{Gorkov:1961}
L.~P.~Gorkov and T.~K.~Melik-Barkhudarov,
Sov.\ Phys.\ JETP {\bf 13}, 1018 (1961).

\bibitem{Baym:2001}
M.~Holzmann, G.~Baym, J.-P.~Blaizot and F.~Laloe,
Phys.\ Rev.\ Lett. {\bf 87}, 120403 (2001).

\bibitem{Petrov:2003}
D.~S.~Petrov, C.~Salomon, G.~V.~Shlyapnikov,
Phys.\ Rev.\ {\bf A 71}, 012708 (2005)
[cond-mat/0407579].
%%CITATION = COND-MAT 0407579;%%

\bibitem{Wingate:2004wm}
M.~Wingate,
%``Exploring lattice methods for cold fermionic atoms,''
preprint, hep-lat/0409060.
%%CITATION = HEP-LAT 0409060;%%

\bibitem{Chang:2004sj}
J.~Carlson, S.-Y.~Chang, V.~R.~Pandharipande, and K.~E.~Schmidt,
%``Superfluid Fermi Gases with Large Scattering Length ,''
Phys.\ Rev.\ Lett.\ {\bf 91}, 050401,  (2003).

\bibitem{Son:1999uk}
D.~T.~Son,
%``Supercond by long-range color mag int in  high-density quark matter,''
Phys.\ Rev.\  {\bf D59}, 094019 (1999)
[hep-ph/9812287].
%%CITATION = HEP-PH 9812287;%%
 
\bibitem{Schafer:1999jg}
T.~Sch{\"a}fer and F.~Wilczek,
%``Superconductivity from perturbative oge in high density  quark matter,''
Phys.\ Rev.\  {\bf D60}, 114033 (1999)
[hep-ph/9906512];
%%CITATION = HEP-PH 9906512;%%
%\bibitem{Pisarski:2000tv}
R.~D.~Pisarski and D.~H.~Rischke,
%``Color superconductivity in weak coupling,''
Phys.\ Rev.\  {\bf D61}, 074017 (2000)
[nucl-th/9910056];
%%CITATION = NUCL-TH 9910056;%%  
%\bibitem{Hong:2000fh}
D.~K.~Hong, V.~A.~Miransky, I.~A.~Shovkovy and L.~C.~Wijewardhana,
%``Schwinger-Dyson approach to color superconductivity in dense QCD,''
Phys.\ Rev.\  {\bf D61}, 056001 (2000)
[hep-ph/9906478];
%%CITATION = HEP-PH 9906478;%%
%\bibitem{Brown:1999aq}
W.~E.~Brown, J.~T.~Liu and H.~c.~Ren,
%``On the perturbative nature of color superconductivity,''
Phys.\ Rev.\ D {\bf 61}, 114012 (2000)
[hep-ph/9908248].
%%CITATION = HEP-PH 9908248;%%

\bibitem{Nickel:2006vf}
D.~Nickel, J.~Wambach and R.~Alkofer,
%``Color-superconductivity in the strong-coupling regime of Landau gauge
%QCD,''
preprint, hep-ph/0603163.
%%CITATION = HEP-PH 0603163;%%

\bibitem{Viverit:2000}
L.~Viverit, C.~J.~Pethick, H.~Smith,
Phys.\ Rev.\ {\bf A61} 053605 (2000) 
[cond-mat/9911080].

\bibitem{Larkin:1964}
A.~I.~Larkin and Yu.~N.~Ovchinikov,
Zh.\ Eksp.\ Theor.\ Fiz.\ {\bf 47}, 1136 (1964);
engl.\ translation:
Sov.\ Phys.\ JETP {\bf 20}, 762 (1965).

\bibitem{Fulde:1964}
P.~Fulde and A.~Ferrell,
Phys.\ Rev.\ {\bf 145}, A550 (1964).

\bibitem{Casalbuoni:2003wh}
R.~Casalbuoni and G.~Nardulli,
%``Inhomogeneous superconductivity in condensed matter and QCD,''
Rev.\ Mod.\ Phys.\  {\bf 76}, 263 (2004)
[hep-ph/0305069].
%%CITATION = HEP-PH 0305069;%%

\bibitem{Alford:1999mk}
M.~Alford, K.~Rajagopal and F.~Wilczek,
%``Color-flavor locking and chiral symmetry breaking in high density {QCD},''
Nucl.\ Phys.\  {\bf B537}, 443 (1999)
[hep-ph/9804403].
%%CITATION = HEP-PH 9804403;%%

\bibitem{Bedaque:2001je}
P.~F.~Bedaque and T.~Sch{\"a}fer,
%``High density quark matter under stress,''
Nucl.\ Phys.\ {\bf A697}, 802 (2002)
[hep-ph/0105150].
%%CITATION = HEP-PH 0105150;%%

\bibitem{Casalbuoni:1999wu}
R.~Casalbuoni and D.~Gatto,
%``Effective theory for color-flavor locking in high density QCD,''
Phys.\ Lett.\ {\bf B464}, 111 (1999)
[hep-ph/9908227].
%%CITATION = HEP-PH 9908227;%%

\bibitem{Kryjevski:2004jw}
A.~Kryjevski and T.~Sch{\"a}fer,
%``An effective theory for baryons in the CFL phase,''
Phys.\ Lett.\ B {\bf 606}, 52 (2005)
[hep-ph/0407329].;
%%CITATION = HEP-PH 0407329;%%
%\bibitem{Kryjevski:2004kt}
A.~Kryjevski and D.~Yamada,
%``CFL phase of high density QCD at non zero strange quark mass,''
Phys.\ Rev.\ D {\bf 71}, 014011 (2005)
[hep-ph/0407350].
%%CITATION = HEP-PH 0407350;%%

\bibitem{Schafer:1998ef}
T.~Sch{\"a}fer and F.~Wilczek,
%``Continuity of quark and hadron matter,''
Phys.\ Rev.\ Lett.\  {\bf 82}, 3956 (1999)
[hep-ph/9811473].
%%CITATION = HEP-PH 9811473;%%

\bibitem{Kryjevski:2005qq}
A.~Kryjevski,
%``Spontaneous superfluid current generation in CFL at nonzero strange quark
%mass,''
preprint, hep-ph/0508180.
%%CITATION = HEP-PH 0508180;%%

\bibitem{Schafer:2005ym}
T.~Sch{\"a}fer,
%``P-wave meson condensation in high density QCD,''
Phys.\ Rev.\ Lett.\ {\bf 96}, 012305 (2006)
[hep-ph/0508190].
%%CITATION = HEP-PH 0508190;%%

\bibitem{Son:2005qx}
D.~T.~Son and M.~A.~Stephanov,
%``Phase Diagram of Cold Polarized Fermi Gas,''
preprint, cond-mat/0507586.
%%CITATION = COND-MAT 0507586;%%

\bibitem{Gerhold:2006dt}
A.~Gerhold and T.~Sch\"afer,
preprint, hep-ph/0603257.
%%CITATION = HEP-PH 0603257;%%

\bibitem{Casalbuoni:2005zp}
R.~Casalbuoni, R.~Gatto, N.~Ippolito, G.~Nardulli and M.~Ruggieri,
%``Ginzburg-Landau approach to the three flavor LOFF phase of QCD,''
Phys.\ Lett.\ B {\bf 627}, 89 (2005)
[hep-ph/0507247];
%%CITATION = HEP-PH 0507247;%%
M.~G.~Alford, J.~A.~Bowers and K.~Rajagopal,
%``Crystalline color superconductivity,''
Phys.\ Rev.\ D {\bf 63}, 074016 (2001)
[hep-ph/0008208].
%%CITATION = HEP-PH 0008208;%%

\bibitem{Ketterle}
M.~W.~Zwierlein, C.~H.~Schunck, A.~Schirotzek, W.~Ketterle,
%``Direct Observation of the Superfluid Phase Transition in Ultracold 
% Fermi Gases''
preprint, cond-mat/0605258.
%%CITATION = COND-MAT 0605258;%%

\bibitem{Thomas}
%``Breakdown of Hydrodynamics in the Radial Breathing Mode of a 
% Strongly-Interacting Fermi Gas''
J.~Kinast, A.~Turlapov, J.~E.~Thomas,
Phys.\ Rev.\ {\bf A 70}, 051401(R) (2004)
[cond-mat/0409283].
%%CITATION = COND-MAT 0409283;%%

\end{thebibliography}
\end{document}